# NetSci High: Bringing Network Science Research to High Schools

Catherine Cramer[1], Lori Sheetz[2], Hiroki Sayama[3,4], Paul Trunfio[5],
H. Eugene Stanley[5], and Stephen Uzzo[1]

[1] New York Hall of Science
[2] Network Science Center, US Military Academy at West Point
[3] Collective Dynamics of Complex Systems Research Group, Binghamton University
[4] Center for Complex Network Research, Northeastern University
[5] Center for Polymer Studies, Boston University
`ccramer@nysci.org, Lori.Sheetz@usma.edu`

**Abstract.** We present NetSci High, our NSF-funded educational outreach program that connects high school students who are underrepresented in STEM (Science Technology Engineering and Mathematics), and their teachers, with regional university research labs and provides them with the opportunity to work with researchers and graduate students on team-based, year-long network science research projects, culminating in a formal presentation at a network science conference. This short paper reports the content and materials that we have developed to date, including lesson plans and tools for introducing high school students and teachers to network science; empirical evaluation data on the effect of participation on students' motivation and interest in pursuing STEM careers; the application of professional development materials for teachers that are intended to encourage them to use network science concepts in their lesson plans and curriculum; promoting district-level interest and engagement; best practices gained from our experiences; and the future goals for this project and its subsequent outgrowth.

## 1 Introduction

Educational systems worldwide are not keeping up with the explosion in the big data and data-driven sciences that inform us about vital trends, have the potential to empower us to solve our greatest social and environmental challenges, and increasingly affect our lives. This gap is coinciding with an escalation in the complexity of the kinds of biomedical, socio-economic, environmental, and technological problems science is addressing, along with the ability to gather and store the subsequent vast amounts of data (American Association for the Advancement of Science (AAAS) 1990, Watts 2007). The skills needed by the 21st century STEM workforce include:

- The ability to interact with large amounts of data. Facility with visual metaphors and granularity for both static and dynamic data streams is needed in order to see patterns in complex data.
- The ability to understand the changing role of models. The higher-order thinking associated with model development allows both exploratory and inductive skills to be used to identify general patterns and characterize their behavior across a wide range of differing environments and processes.

Students in the STEM "pipeline" need to be prepared for this new reality as they enter the modern day workforce and higher education. However, exposure to these data-driven science skills is unavailable to most primary and secondary school students. Furthermore, summer or academic year research experiences for high school students under researchers' guidance are often inaccessible to disadvantaged young learners. Such lack of access sends students down a path that is devoid of opportunities to fully participate in advances in our modern society.

Network science has emerged as a possible solution. It is a promising way to address data-intensive real-world problems, employing graph theory, statistical analysis and dynamical systems theory to large, complex data sets, seeking patterns and leveraging them against large-scale knowledge management and discovery in business, medicine, policymaking, and virtually all complex science disciplines today. Network science is being used to understand everything from the human brain, to the origins of cancer, to the growth of cities, to our impact on the environment (Barabasi 2002, Pastor-Satorras and Vespignani 2001, Lazer et al 2013). Network science demands that we revise our thinking about what kinds of technical and process skills are needed to design, create and explore these emerging and accumulating data and technological structures.

We believe that network science can provide a novel pathway for high school students to learn about traditional topics across many disciplines, including social studies, science, computer science and technology. Many of the problems explored through a network science approach are in the everyday experience of students, such as the network flow of air traffic, interconnectivity of coupled networks in political and social systems, and human networks as seen through technology activities such as Facebook and Twitter.

To test this solution – using network science to close the skills gap – we have developed and are running "NetSci High", a regional educational outreach program designed to empower high school students and teachers to harness the power of network modeling and analysis, resulting in a more holistic, dynamic understanding of the "interdependence" among components and the evolution of relationships among various things around us. NetSci High provides interventions in STEM teaching and learning that directly address the need for twenty-first century skills while targeting female, minority and economically disadvantaged students. It provides an advanced, alternative pathway to developing rigorous skills-based curricula, resources and programs that utilize the rapidly growing science of complex networks as a vehicle through which students can learn computational and analytical skills for network-oriented data analysis, as well as how these skills can lead to breakthroughs in solving large-scale, real-world problems. NetSci High explores innovative approaches that, as our work demonstrates, can capture the interest and imagination of underrepresented populations to explore science research problems using computational tools and methods (Buldyrev et al. 1994, Cohen et al. 2000, Trunfio et al. 2003).

## 2 What Is NetSci High?

NetSci High began as a small pilot project in 2010 with financial support from the Office of International Science and Engineering at the US National Science Foundation (NSF) as well as a corporate donation from BAE Systems. The first year of NetSci High (2010-2011) was run as a competition for high school research posters. Seven student projects were conducted through collaboration between participating high schools and their local research labs in New York City, Boston, and Binghamton, NY. Their posters were reviewed by the Scientific Committee, and the students and teachers of the two winning posters were supported to attend the NetSci 2011 conference in Budapest, Hungary, in June 2011. All of the seven posters were presented at the poster sessions of the conference. The posters were also displayed at the Eighth International Conference on Complex Systems in Boston on June 26-July 1, 2011. The second year of NetSci High (2011-2012) was run as scholarships offered to high school student teams. Two student teams participated from the Binghamton area. Those teams were offered a scholarship to attend the NetSci 2012 conference in Evanston, IL, in June 2012, and to present their posters.

This pilot program paved the way to a much larger NSF-funded ITEST Strategies project, "Network Science for the Next Generation," or NetSci High. Since 2012, Boston University, the New York Hall of Science, SUNY Binghamton and United States Military Academy at West Point have been collaborating on this ITEST Strategies project, which provides opportunities for disadvantaged high school students to participate in cutting-edge network science research. This project bridges information technology practice and advances in network science research to provide career and technical education opportunities for young people underrepresented in data-driven STEM.

The goals of NetSci High are the following:

1. Improve computational and statistical thinking and stimulate interest in computer programming and computational scientific methods by providing students and teachers with opportunities to create and analyze network models for real-world problems through a mentoring and training program.
2. Increase students' potential for success in STEM in a technical career or college through applied problem solving across the curriculum using tested units of instruction that clarify complex STEM topics and provide new applied approaches for critical thinking in STEM.
3. Prepare learners for $21^{st}$ century science and engineering careers through the use of data-driven science literacy skills, and motivate them to elucidate social and scientific problems relevant to the disciplines and to their lives.
4. Develop curricular resources that help learners achieve the following set of basic skills that are crucially needed to succeed in the data-driven work environment in the $21^{st}$ century:

    — *Ability to synthesize*, seek and analyze patterns in large-scale data systems;
    — *Gain facility with data visualization*, filtering, federating, and seeking patterns in complex data;

- *Understand the changing role of models,* higher-order thinking, emphasizing exploratory skills to identify and characterize behavior of patterns in differing environments;
- *Use network science and statistical approaches* to break down traditional silos in order to compare and contrast processes across domains;
- *Build data fluency* to be able to identify, clean, parse, process and apply appropriate analysis skills to large quantities of data;
- *Gain facility with data mining and manipulation* with increasingly semantic and statistical approaches, superseding logic models for searching and comparing data; and
- *Understand the role of data sharing*, collaboration, interoperability of tools and data types, along with skills in using collaborative tools and methods to maximize data discovery.

## 3    Contents and Materials

NetSci High has developed and implemented a rich, experiential, research-based program for disadvantaged high school students, science research graduate student mentors, and high school STEM teacher mentors throughout New York State and Boston, Massachusetts. This project works to close the gap between the teaching and learning of STEM disciplines and STEM practice, and to prepare the next generation of the STEM workforce to conduct a mode of research that differs markedly from that currently mandated by K-12 curricula and educational practice.

The program includes a 2-week intensive summer workshop and an academic year research program utilizing collaborative IT tools, plus periodic special workshops, industry lab tours, and participation in the International School and Conference on Network Science (NetSci).

Organizers have assembled teams of high school students from New York State and Boston area Title 1 schools, plus their science teachers and graduate students from network science research labs, to spend a year collaborating on cutting-edge research on a network science topic of their choosing. The research component of this project focuses on the efficacy of intensive training and support of high school student teams and an academic year of research with cooperating university-based network science research labs; the labs' participation is facilitated by a graduate student who learns valuable mentorship skills as part of the experience. Because the network science field is relatively new, much of this research is novel, with practical implications.

Each yearly cohort of students begins their experience by participating in a summer residency-workshop led by network science faculty and researchers (Fig. 1). This summer workshop is an intensive two-week experience, at which student teams, their teachers and graduate student mentors are immersed in learning network science concepts and programming languages such as Python, NetLogo and JavaScript; applying network analysis tools such as NetworkX and Gephi; attending hands-on workshops and talks given by top network science researchers; and collaboratively brainstorming about research questions that will form the basis of the year's research projects.

During the academic year students refine their coding and programming skills, conduct their research, visit their host research labs, and have regular weekly meetings with graduate student mentors. Layers of support and mentoring throughout the academic year come from graduate students in partnering labs, high school teachers and project staff, as well as online and face-to-face field trips, meetings, seminars and work sessions. After their research is complete the teams prepare their findings for publication and presentation.

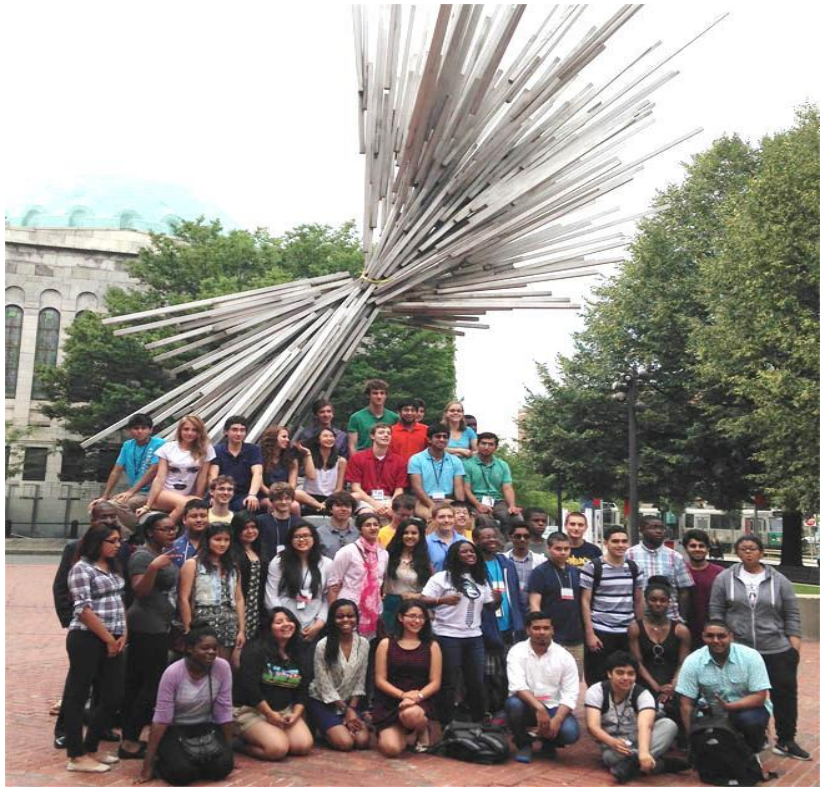

**Fig. 1.** NetSci High New York teams gather for a conference at the close of the July 2014 workshop at Boston University.

Evaluation of this model to date has indicated that it is extremely effective for students in incentivizing and achieving success in mastering and applying data-driven STEM skills to real problem solving. It is an empowering and engaging pathway into data and computational literacy and computer programming skills. Further planned evaluation will aim to assess participants' higher education and career choices and their relevance to STEM fields. The subsequent evaluative results will provide substantive evidence of the return on investment (ROI) for funding priorities and future growth of the project.

More information can be found at the NetSci High websites (https://sites.google. com/a/nyscience.org/netscihigh/ and http://www.bu.edu/networks/). Professional development and workshop resources including curricular modules can be found at the workshop website (http://www.bu.edu/networks/workshop/). Additional resources for network science education can be found on the NetSciEd website (https://sites. google. com/a/nyscience.org/netscied/).

## 4   Accomplishments

Over the past four years, high school students have worked on a wide variety of research projects through the NetSci High program. Table 1 shows the list of project titles, listed in chronological order.

**Table 1.** List of titles of NetSci High student research projects

2010-11
- A Comparative Study on the Social Networks of Fictional Characters
- Academic Achievement and Personal Satisfaction in High School Social Networks
- Does Facebook Friendship Reflect Real Friendship?
- Inter-Species Protein-Protein Interaction Network Reveals Protein Interfaces for Conserved Function
- The Hierarchy of Endothelial Cell Phenotypes
- Preaching To The Choir? Using Social Networks to Measure the Success of a Message
- Identification of mRNA Target Sites for siRNA Mediated VAMP Protein Knockdown in *Rattus Norvegicus*

2011-12
- A Possible Spread of Academic Success in a High School Social Network: A Two-Year Study
- Research on Social Network Analysis from a Younger Generation

2012-13
- Interactive Simulations and Games for Teaching about Networks
- Mapping Protein Networks in Three Dimensions
- Main and North Campus: Are We Really Connected?
- High School Communication: Electronic or Face-to-Face?
- An Analysis of the Networks of Product Creation and Trading in the Virtual Economy of Team Fortress 2

2013-14
- A Network Analysis of Foreign Aid Based on Bias of Political Ideologies
- Comparing Two Human Disease Networks: Gene-Based and System-Based Perspectives
- How Does One Become Successful on Reddit.com?
- Influence at the 1787 Constitutional Convention

- Quantifying Similarity of Benign and Oncogenic Viral Proteins Using Amino Acid Sequence
- Quantification of Character and Plot in Contemporary Fiction
- RedNet: A Different Perspective of Reddit
- Tracking Tweets for the Superbowl

NetSci High has facilitated sending a group of high school students and teachers from New York City to NetSci 2011 in Budapest, Hungary; a group from Endwell and Vestal, NY to NetSci 2012 in Evanston, IL; and a group from Vestal, NY to NetSci 2014 in Berkeley, CA. The high school student teams presented their work at poster sessions at all of these conferences. High school student research has also been published in peer-reviewed journals such as PLOS ONE (Blansky et al. 2013).

During the Spring 2014 semester, professional development training in network science concepts and tools was provided to the entire 9th grade faculty at Chelsea CTE High School in New York City and faculty from Newburgh Free Academy. The faculty were then encouraged to apply these concepts to meet objectives in their current lesson plans.

Historically each of the program elements have been initiated by network science researchers approaching school district administrators and teachers with ideas for research programs, professional development, workshops, and events all to bring network science concepts, tools and resources into the high school as an apparatus for learning. While the districts appreciated the successes realized by the participating students and teachers, the districts had not developed a deep enough understanding of network science to grasp the full potential of using network science as a curriculum tool. In Fall 2014 district administrators reached out to mentor teachers requesting development of a high school level network science elective course. This request represents a fundamental shift at the district level. The districts are now reaching out to the network science community seeking more resources to bring to their students.

To more widely disseminate the success of this program and to meet the demand for expanding the role of network science in education (a demand coming both from the high school educational community as well as the community of network scientists), the partners initiated a symposium at NetSci 2012 called Network Science in Education (NetSciEd). Since then these symposia have been held annually in the U.S. and Europe and have led to a significant rise in interest in educational and learning applications of network science and the subsequent formation of the NetSciEd community. The NetSciEd community undertook the development and articulation of a set of Network Literacy Essential Concepts that all citizens should know by the time they graduate from high school. These can be found on the NetSciEd website (https://sites.google.com/a/nyscience.org/netscied/). Moreover, for the first time, Networks and Education will be an official strand at the 2015 International NetSci Conference to be held in Zaragoza, Spain, in June 2015 (http://netsci2015.net/).

## 5      Conclusions and Future Work

As described above, NetSci High has made significant educational impacts on regional high school students and teachers, and is also prompting strong social commitments from the Network Science community as a whole (Harrington et al 2013, Sanchez and Brandle 2014). It aims to address the challenge of transforming the way we educate our citizens in order to keep pace with not only the amount of data we collect, but to appreciate how network science identifies, clarifies, and solves complex 21st century challenges in the environment, medicine, agriculture-urbanization, social justice and human wellbeing. This project provides a pathway to integrate science research and programming skills for high school students who would not otherwise have these opportunities. Additionally, this project encourages high school teacher mentors to broaden their STEM understanding and informs their current teaching in terms of content and practice.

Through evaluation and remediation in our current NetSci High project, we have identified the following successful strategies for bringing network science into high school teaching and learning:

- Original student and teacher research projects are not only possible, but form an essential incentive and commitment for participants to remain engaged in and to bring projects to completion.
- While there is significant interest in broad collaboration among teachers in different domains, they prefer to start with small, easily definable curriculum units or lessons that can be implemented within a class.
- It is possible to train a broad spectrum of students and teachers in enough computer programming (e.g., Python or R) to use sophisticated network analysis tools within programming environments.
- A supportive community and consistent mentorship are essential to success.
- Teachers can and have assumed an active leadership role in mentoring students who are engaged in network science research, provided the right supports are in place.
- Students and teachers are remarkably innovative in terms of how they develop and pursue project-based learning approaches in network science.
- It became immediately apparent in the first year of the project that participating teachers were most effective if they had the same level of training as their students: they want to be active mentors, rather than co-learners side by side. As the project progressed and interactions with teachers became deeper and more meaningful and teachers took on new roles and pursued their own interests and took ownership of network science approaches, a path to scalability began to emerge.

NetSci High is still developing, and there are a number of aspects on which further development and expansions are needed. To accomplish this, we are looking for support and collaboration from the entire Network Science community. Over the next few years, we plan to achieve the following in order to make this successful program more organized, more scalable and more accessible to everyone on the globe:

- Refine our learning materials, publish a Network Science Workshop Training Manual, and develop network science mobile teaching kits. Such field resources will be a blueprint for future training of participants as well as disseminating and replicating our work.
- Expand on successful live network science professional development workshops for high school teachers and develop interdisciplinary network science curriculum.
- Promote new projects on data mining of educational data and using network science to understand the performance of educational institutions. There is an increasing amount of work looking at the organizational structure of education through a network science lens, particularly at how we might mine student data and use network analysis to determine the impact of churn on organizational structure and efficaciousness in schools and districts.
- Finalize the Network Literacy Essential Concepts that all citizens should know. This will be used to create a framework for developing curriculum that can better support data-driven STEM than is currently possible, and will support Next Generation Science Standards (NGSS Lead States 2013) and Common Core standards (National Governors Association Center for Best Practices 2010).
- Expand the professional development, student research and curriculum development projects that benefit from a global community of scientists and policymakers who see network science as an accessible entry point for vital computational, data literacy and algorithmic skills.
- Maintain and increase dialog with the private sector to expand support for initiatives in network science in teaching and learning and engage STEM professionals in awareness and participation in this work.
- Author an accessible network science e-book for general readership.
- Establish a network science e-badging system for the entire network science teaching, learning and research community.
- Host a network and data science festival for the public.
- Develop international partnerships with network science researchers and educators outside the US and promote similar educational activities at international scales.

## Acknowledgments

The NetSci High program is supported by the US National Science Foundation through the Cyber-enabled Discovery and Innovation program (CDI) and the Office of International Science and Engineering (OISE) (Award # 1027752) and the Innovative Technology Experiences for Students and Teachers program (ITEST) (Award # 1139478/1139482), as well as a corporate donation from BAE Systems.